\documentclass[twocolumn,twocolappendix]{aastex631}

\shorttitle{SALT3 Training Systematics}
\shortauthors{Dai et al.}

\graphicspath{{./}}
\usepackage{xcolor}
\usepackage{float}
\usepackage{amsmath}
\usepackage{hyperref}
\usepackage{tabularx}
\usepackage{chngcntr}

\defcitealias{Kenworthy21}{K21}
\newcommand\numberthis{\addtocounter{equation}{1}\tag{\theequation}}

\newcommand{\nolowz}{NO-LOWZ}
\newcommand{\nou}{NO-U}
\newcommand{\miscal}{MIS-CAL-SPEC}
\newcommand{\halfspec}{HALF-SPEC}
\newcommand{\hosthundred}{HOST-100}
\newcommand{\hostfifty}{HOST-50}
\newcommand{\hostten}{HOST-10}
\newcommand{\byobase}{BYO-STRETCH-COLOR}
\newcommand{\byovel}{BYO-VEL}
\newcommand{\byohost}{BYO-HOST}
\newcommand{\byohostz}{BYO-HOST-Z-DEP}

\begin{document}

\title{Propagating Uncertainties in the SALT3 Model Training Process to Cosmological Constraints}

\correspondingauthor{M.~Dai}
\email{mi.dai@jhu.edu}

\author[0000-0002-5995-9692]{M.~Dai}
\affiliation{Department of Physics and Astronomy, The Johns Hopkins University, Baltimore, MD 21218, USA}

\author[0000-0002-6230-0151]{D.~O.~Jones}
\affil{Gemini Observatory, NSF's NOIRLab, 670 N. A'ohoku Place, Hilo, Hawai'i, 96720, USA}

\author[0000-0002-5153-5983]{W.~D.~Kenworthy}
\affil{The Oskar Klein Centre for Cosmoparticle Physics, Department of Physics,Stockholm University, SE-10691 Stockholm, Sweden.}

\author[0000-0003-3221-0419]{R.~Kessler}
\affil{Kavli Institute for Cosmological Physics, University of Chicago, Chicago, IL 60637, USA} 
\affil{Department of Astronomy and Astrophysics, University of Chicago, 5640 South Ellis Avenue, Chicago, IL 60637, USA}

\author[0000-0002-2361-7201]{J.~D.~R.~Pierel} 
\affil{Space Telescope Science Institute, Baltimore, MD 21218, USA}

\author[0000-0002-2445-5275]{R.~J.~Foley}
\affil{Department of Astronomy and Astrophysics, University of California, Santa Cruz, CA 95064, USA}

\author[0000-0001-8738-6011]{S.~W.~Jha}
\affil{Department of Physics \& Astronomy, Rutgers, State University of New Jersey, 136 Frelinghuysen Road, Piscataway, NJ 08854, USA}

\author[0000-0002-4934-5849]{D.~M.~Scolnic}
\affil{Department of Physics, Duke University, Durham, NC 27708, USA}

\begin{abstract}
Type Ia supernovae (SNe\,Ia) are standardizable candles that must be modeled empirically to yield cosmological constraints. To understand the robustness of this modeling to variations in the model training procedure, we build an end-to-end pipeline to test the recently developed SALT3 model.  We explore the consequences of removing pre-2000s low-$z$ or poorly calibrated $U$-band data, adjusting the amount and fidelity of SN\,Ia spectra, and using a model-independent framework to simulate the training data.  We find the SALT3 model surfaces are improved by having additional spectra and $U$-band data, and can be shifted by $\sim$5\% if host galaxy contamination is not sufficiently removed from SN spectra.   We find that resulting measurements of $w$ are consistent to within 2.5\% for all training variants explored in this work, with the largest shifts coming from variants that add color-dependent calibration offsets or host galaxy contamination to the training spectra, and those that remove pre-2000s low-$z$ data. These results demonstrate that the SALT3 model training procedure is largely robust to reasonable variations in the training data, but that additional attention must be paid to the treatment of spectroscopic data in the training process.  We also find that the training procedure is sensitive to the color distributions of the input data; the resulting $w$ measurement can be biased by $\sim2\%$ if the color distribution is not sufficiently wide. Future low-$z$ data, particularly $u$-band observations and high signal-to-noise ratio SN\,Ia spectra, will help to significantly improve SN\,Ia modeling in the coming years.

\end{abstract}


\section{Introduction} \label{sec:intro}

Since the discovery of cosmic acceleration \citep{Riess98,Perlmutter99}, Type Ia supernovae (SNe\,Ia) have played an important role in constraining the dark energy equation of state, $w$. Calibrated SN\,Ia distances at low redshift are also used to derive the most precise local measurements of the Hubble constant (H$_0$; \citealt{Riess22}), currently in tension with inferred H$_0$ values from the cosmic microwave background (\citealp{Planck18}; see reviews from \citealp{Verde19,DiValentino21}).

The most recent cosmological constraints from SNe\,Ia use up to $\sim$1500 SNe Ia \citep{Scolnic18,Jones19,Brout22} and in the near future, surveys such as Vera Rubin Observatory's Legacy Survey of Space and Time \citep{LSSTDESC18} and the SN survey from the {\it Nancy Grace Roman Telescope} \citep{Hounsell18, Rose21b} will increase the number of well-measured SNe\,Ia by orders of magnitude. Although the statistical uncertainties of the cosmological measurements will be greatly reduced with these future data sets, understanding and reducing the systematic uncertainties will be crucial.  Most previous studies have found that the dominant systematic uncertainties in SN\,Ia cosmology analyses are caused by photometric calibration of the data in the cosmology sample and the sample used to train the SN for standardization
\citep{Betoule14,Scolnic18,Brout19,Jones19,Brout22}.  However, explorations of systematic uncertainties in the SN standardization model are typically
{\it limited} to photometric calibration offsets; potential systematic errors
in the training process or definition of the model have been explored \citep[e.g.,][]{Mosher14} but are
rarely propagated to cosmological parameter uncertainty budgets.

Substantial recent effort has been put into developing new SN Ia models \citep{Saunders18,Leget20,Boone21,Pierel21,Kenworthy21, Mandel22}. All SN\,Ia models for cosmology to date are empirical models and therefore require a training sample of well-measured SNe (with or without spectral data); therefore, it is important to understand how robust a SN model is to the choice and quality of the training sample, and the effect of that training sample on the resulting cosmological parameter measurements.  In particular, given the large impact of photometric calibration uncertainties on the SN model training, other systematic uncertainties related to the model training must also be better understood.
Finally, it is important to understand the ways in which new data could enhance the model,
such as data from ATLAS \citep{Tonry18}, the Zwicky Transient Facility \citep{Bellm19}, the Carnegie Supernova Project \citep[CSP,][]{Krisciunas17}, and the Young Supernova Experiment \citep{Jones21}.

The SALT2 model \citep{Guy07,Guy10,Betoule14,Taylor21} is the baseline SN standardization model used for nearly all measurements of $w$ in recent years.  Advantages of the SALT2 model include its ability to use high redshift SNe for training, built-in $k$-corrections due to its continuous SED model across both phase and wavelength, and the fact that the SN amplitude is a free parameter in the training procedure, making the training independent of the cosmological model.  Unfortunately, the original training code and data are not fully publicly available, and are difficult to modify and improve for systematic studies, with \citet{Mosher14} being the most recent study to explore alterations to the SALT2 training procedure.  However, a new ``SALT3" model has recently been developed \citep[hereafter \citetalias{Kenworthy21}]{Kenworthy21}, which shares much of the functional form as SALT2 but includes improvements in the training procedure and includes new data that extends the model wavelength range from $\sim 0.9$ to $1.1 \  \mathrm{\mu m}$.  Though we use the \citetalias{Kenworthy21} model in this work, we note that \citet{Pierel22} recently extended the SALT3 model surfaces to 2~$\mu{\rm m}$, albeit using a data set with somewhat larger calibration uncertainties, and \citet{Jones22} built a SALT3 model that includes a host-galaxy dependent model surface.

Here, we use the open-source SALT3 model training code, {\tt SALTshaker}\footnote{\url{https://github.com/djones1040/SALTShaker}}, to quantify how training data variations or unknown physics in the data can affect the measurement of cosmological parameters.  In particular, we investigate the exclusion of low-$z$ training data without precisely measured filter throughputs, the exclusion of poorly calibrated ground-based near-UV bands, variations in the number of SN\,Ia spectra,
and the impact of wavelength-dependent calibration errors and host-galaxy contamination in those spectra using simulations generated from a previously trained SALT2 model \citep{Pierel18}.  Furthermore, 
we use simulations generated from BYOSED \citep{Pierel21}, a model framework that is independent of SALT, to determine the sensitivity of cosmological parameters to SED features that are not modeled by the SALT framework. 
The BYOSED framework generates SN Ia SEDs using a base model with added ``perturbers" derived from composite spectra of different SN Ia sub-populations; this produces variations in the SN spectra that are not modeled with existing frameworks, such as velocity and host-galaxy mass variations.

For each of these training scenarios, we use a simulation-based approach to quantify how variations in the available training data or unknown physics affects the recovery of the trained model, the correlations between light-curve parameters and luminosity, and the measured value of $w$.  
For this purpose, we have built an end-to-end simulation and analysis pipeline, which will enable future analyses to propagate modeling systematics into the systematic uncertainty budget of cosmological constraints.

In Section \ref{sec:pipeline} we describe our analysis and pipeline, and in Section \ref{sec:data} we discuss our simulation approaches and training sample variations. In Section \ref{sec:results} we discuss the recovered models, nuisance parameters, and cosmological parameters, and in Section \ref{sec:discussion} we discuss our results and conclude.

\section{Description of the analysis and the pipeline}
\label{sec:pipeline}

Modern SN\,Ia cosmology analyses include a complex set of stages to go from input photometric SN\,Ia light curve data to cosmological parameter measurements.  Those stages include fitting for light-curve parameters in order to standardize the SN brightness, correcting for observational biases using simulations, estimating nuisance parameters that relate stretch and color to brightness, computing a covariance matrix to account for systematic uncertainties, and finally, fitting for cosmological parameters.  The community has developed tools to perform these individual tasks \citep[e.g.,][]{Kessler09b,Guy10,Rubin15,Kessler17} and a pipeline to control the workflow \citep{Hinton20}.
 
 Typically missing from the evaluation of systematic uncertainties, however, are estimations of systematics related to the training procedure.
 Here, we build an end-to-end cosmological analysis pipeline that includes a SN model training stage.  While previous studies such as the Pantheon$+$ analysis \citep{Scolnic21} have performed model re-training to incorporate calibration offsets in the SALT model, here we build a more flexible framework for SALT training options and systematic uncertainty evaluation.
This open-source pipeline is built in {\tt Python} and ties together pre-existing code to perform many of the steps necessary to go from input data or simulations to cosmological parameter estimation.  In particular, many stages are built around methods implemented within the SNANA software \citep{Kessler10,Kessler19}; SNANA is a collection of SN analysis methods that perform simulations, light curve parameter estimation, distance measurement, and cosmological parameter fitting. The individual components of the pipeline are described in more detail below.

 The input model in this analysis pipeline is flexible, but the model training and analysis stages currently assume a SALT3 model defined as 
 \begin{align} \label{eq:salt3spectralfluxmodel}
    F(p,\lambda) = &x_0 [M_0(p,\lambda;\boldsymbol{m_0}) + x_1 M_1(p,\lambda;\boldsymbol{m_1})] \nonumber\\ & \cdot \exp(c \cdot CL(\lambda;\boldsymbol{cl})),
\end{align}
where $M_0$ and $M_1$ are model components that describe an average spectral surface and its first-order variation, and $CL$ is the color law, which is a polynomial between 2800 and 8000\AA\ and is extrapolated beyond those wavelength ranges. The parameters $x_0$, $x_1$ and $c$ are light curve parameters that describe the overall amplitude, shape, and color of the light curve.
 
Using those light-curve parameters, distances are measured using the Tripp estimator \citep{Tripp98}:
 
 \begin{equation}
     \mu = m_B + \alpha \times x_1 - \beta \times c + \mathcal{M} + \Delta\mu_{\mathrm{bias}},
\label{eqn:tripp}
 \end{equation}
 
 \noindent where $x_1$ and $c$ are stretch and color parameters from a SALT3 light-curve fit, and $m_B$ is the log of the light-curve amplitude $x_0$.  Nuisance parameters $\alpha$ and $\beta$ are estimated 
 to determine the correlation of $x_1$ and $c$ with luminosity.  $\mathcal{M}$ is a combination of the SN absolute magnitude and H$_0$, which are degenerate in this work. $\Delta\mu_{\mathrm{bias}}$ is the bias correction term that is determined from simulations and used to correct for selection biases. Most versions of the Tripp estimator used in the last decade also include the variable $\gamma$, the correlation between host-galaxy mass and distance measurement \citep{Kelly10,Lampeitl10,Sullivan10}, but we do not include it in this work since it is a second-order effect and requires proper simulation of the relations between the host-galaxy mass, the SN light-curve parameters, and the SN brightness \citep{Smith20b,Popovic21a}.

\begin{figure*}[htb!]
\plotone{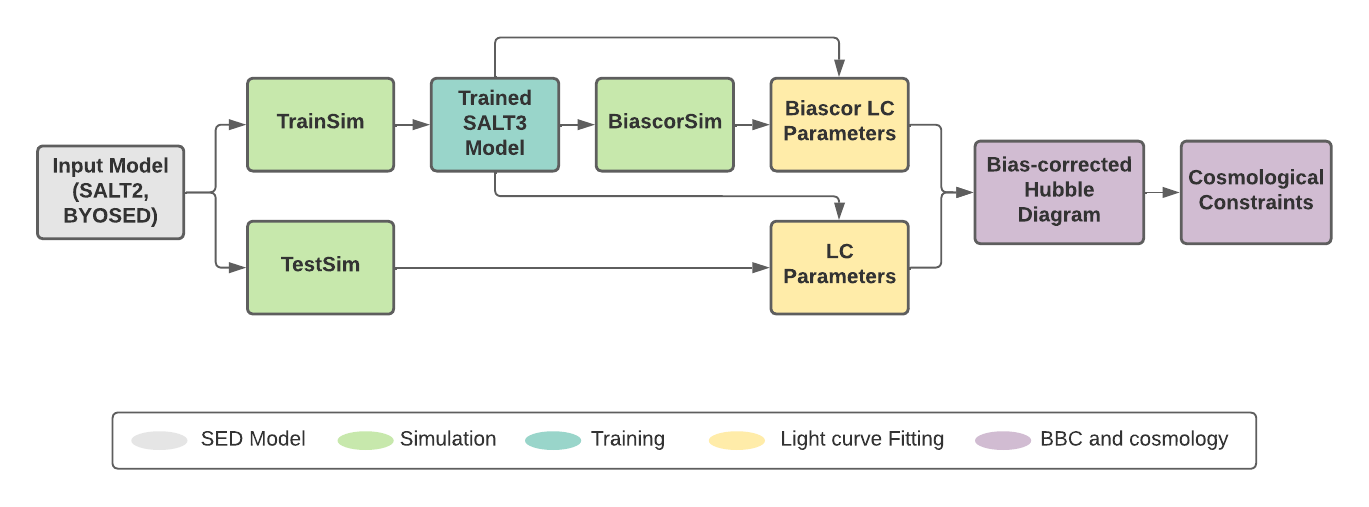}
\caption{Schematic overview of our analysis pipeline, from simulation of the underlying SED model to cosmological parameter constraints. The arrows show the flow of the pipeline and how each stage is connected.
\label{salt3_pipeline}}
\end{figure*}

The stages of the pipeline are described below and illustrated in Figure \ref{salt3_pipeline}:

\begin{itemize}
    \item {\bf Generating Simulated Input Data}.
    The pipeline uses the SNANA simulation program to generate ``observed'' light curves and spectra from a specific SED model.  
    The simulation uses cadences, zeropoints, seeing, host-galaxy noise, and detection criteria from real surveys, along with intrinsic SN parameter distributions and scatter models, to yield realizations of the input model. 
    Here, to allow some degree of SALT3 independence in the evaluation of {\tt SALTshaker}, we use the extended SALT2 model from \cite{Pierel18} as our input model; ``SALT2-extended" is an ultraviolet and near-infrared extrapolation of SALT2.4 \citep{Betoule14} based on low-$z$ SN photometry and spectroscopy. We also generate simulations using the BYOSED model \citep[see Section \ref{subsec:byosed_variants} for details]{Pierel21} to test the robustness of the SALT3 training procedure to effects that are not explicitly modeled by the SALT framework. 
    The pipeline generates both a ``training" simulation for model training and a ``test" sample for cosmological parameter measurement, though in practice --- and in the case of real data --- these samples overlap.
    \item {\bf SN Model Training}.  The training simulations are used as input to the {\tt SALTshaker} code \citepalias{Kenworthy21}, which uses those data to train a new version of the SALT3 model.  The simulated light curves of the training sample are first fit using the SALT2-extended model to provide initial estimates of the time at peak and the light curve parameters as input to the training process.
    \item {\bf Generating ``BiasCor" Simulations}.   To correct the measured light curve parameters for sample selection biases, the pipeline again generates a large sample of SNe\,Ia from the newly trained SALT3 model; this sample is typically a factor of 10-100 times larger than the training or test samples to prevent noise in the bias corrections from significantly contributing to distance uncertainties. The extent to which SALT3 does not accurately model the true SN\,Ia features will propagate to errors in the bias correction. 
    \item {\bf Light-Curve Fitting}. 
    The simulated photometry of the test sample and the BiasCor sample are fit with the newly trained SALT3 model.
    \item {\bf Bias Correction and Distance Measurement}.  After light-curve fitting, the BiasCor simulations are used to determine a mapping between simulated versus measured light-curve parameters as a function of redshift using the ``BEAMS with Bias Corrections (BBC)" method \citep{Kessler17}. 
    The BBC method determines the nuisance parameters $\alpha$ and $\beta$ and applies Equation \ref{eqn:tripp} to measure distances.
    For computational efficiency, we use a 1D bias correction method within the BBC framework, which corrects those distances as a function of redshift for selection effects and returns maximum-likelihood distances binned by redshift.   
    \item {\bf Cosmological Parameter Estimation}.  Finally, the pipeline uses these distances to fit a $w$CDM model in a maximum likelihood fit to measure $w$ with a cosmic microwave background (CMB) prior using the $R(z_*)$ shift parameter (e.g. Eq. 69 in \citealt{Komatsu09}) with $\sigma_R=0.007$.  This is a computationally efficient way of producing CMB constraints with a constraining power that is similar to \citet{Planck18}.  Finally, the pipeline
    evaluates  biases in the measured cosmology with respect to the input cosmology.
\end{itemize}

Our pipeline is publicly available from the  {\tt SALTShaker} package\footnote{\url{https://saltshaker.readthedocs.io/en/latest/pipeline.html}.}.

\section{Evaluating Systematic Uncertainties from the SALT Model}
\label{sec:data}

To quantify biases on measurements of cosmological parameters due to the SALT3 training procedure, we create realistic simulations and use them as input ``data'' to the {\tt SALTShaker} training code. In different simulations, we vary the training set or the underlying properties of the SED model used to generate the SN\,Ia sample.  We describe each of these simulations in detail below and we analyze the effect of each simulation on the trained SALT3 model and on the recovered cosmological parameters after re-training the SALT3 model for each in Section \ref{sec:results}.

\subsection{Baseline simulation}
\label{subsec:baseline_sim}

Using SALT2-extended, we first create a baseline simulation that mimics the SALT3 training set from \citetalias{Kenworthy21}, including both photometry and spectra. The SALT3 training sample includes a low-$z$ compilation --- the Calan/Tololo survey \citep{Hamuy96}, CfA1-4 \citep{Riess99,Jha06,Hicken09,Hicken12}, the Carnegie Supernova Project (CSP; \citealp{Krisciunas17}), and the Foundation Supernova Survey \citep{Foley18,Jones19} --- as well as higher-$z$ data from SDSS \citep{Holtzman08}, SNLS \citep{Astier06}, PanSTARRS \citep{Rest14,Jones18,Scolnic18}, and DES \citep{DES18}. The sample size varies for each random realization; on average, each random sample includes a total of 1047 SNe with 1168 spectra, slightly smaller than the SALT3.K21 training set (1083 SNe and 1207 spectra) due to random selections in the simulation process.

In \citetalias{Kenworthy21}, SNANA simulations were generated that reproduced the measured light-curve parameters of the real data in order to create a simple approximation of the distribution of data for a simulation-based training.  Here, we instead adopt an approach in which light-curve parameters are drawn from Monte Carlo-sampled distributions that will on average yield the same $x_1$ and $c$ distributions as the real data.  %
In the baseline simulations and the variants in Section \ref{subsec:trainingvariants} below, we use SALT2-extended as our input model to allow our simulated model to be independent of the {\tt SALTShaker} training process.

\begin{figure}[htb!]
\centering
\includegraphics[trim=1cm 0.1cm 0.2cm 0.1cm, width=0.45\textwidth]{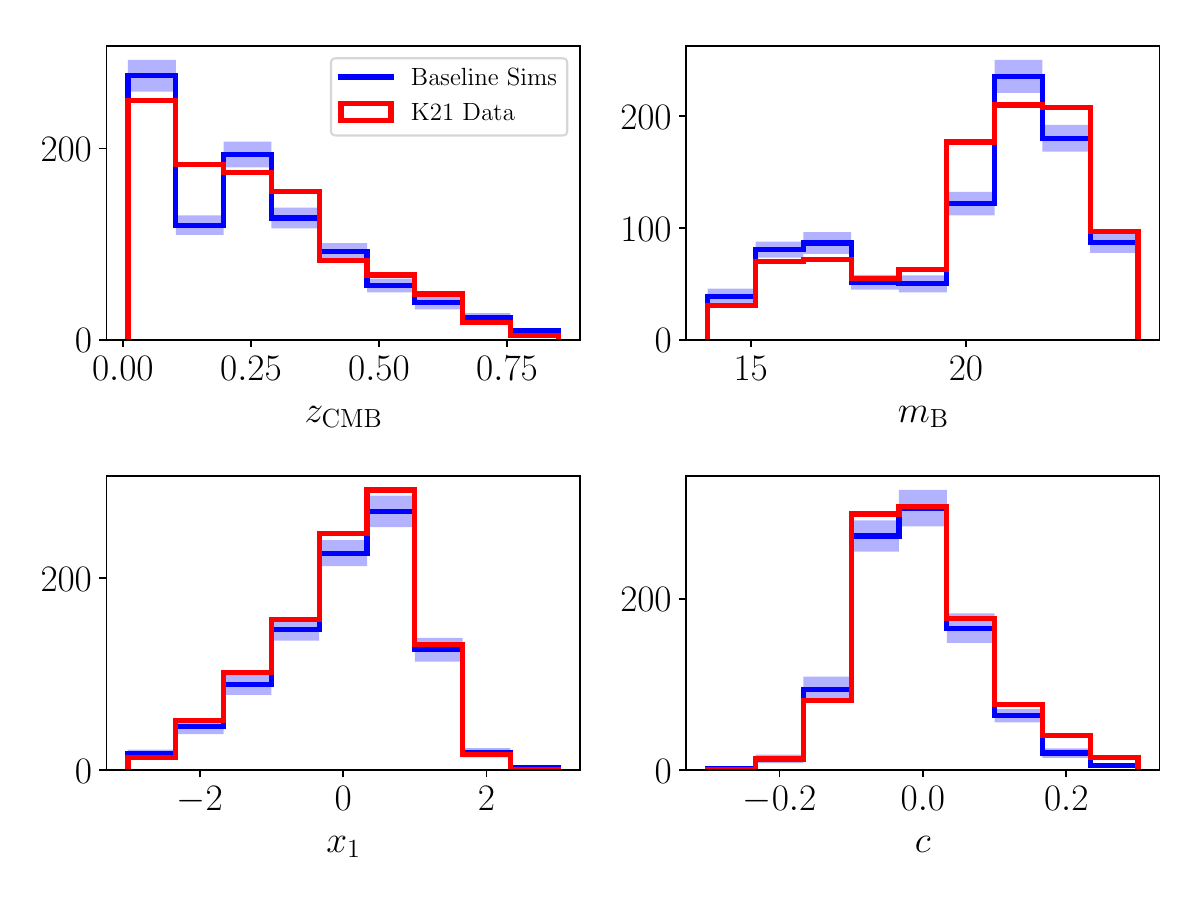}
\caption{Comparison of light curve parameter distributions between the average of 40 baseline simulations (Section \ref{subsec:baseline_sim}) and real \citetalias{Kenworthy21} data. Blue histograms are the simulated baseline data, with the shaded area being the standard deviation of the number in each bin from the simulations, and red histograms are the \citetalias{Kenworthy21} data.
\label{fig:lcpar}}
\end{figure}

For the samples discussed above, SNANA simulations have already been developed as part of previous cosmological analyses.  The low-$z$, SDSS, Pan-STARRS and SNLS simulations used here are from \citet{Scolnic18} with $x_1$ and $c$ populations from \citep{Scolnic16}, the DES simulations are from \citet{Kessler19b}, and the Foundation simulations are from \citet{Jones19}. We %
keep the same $x_1$/$c$ distributions that are developed by the works above. The parameters of the $x_1$/$c$ distributions for each survey are summarized in Appendix \ref{ap:sim_pars}. We note that the same distributions are used for simulating both the training sample (TrainSim) and the test sample (TestSim) in the pipeline. Each pipeline run generates one TrainSim for one TestSim. A comparison between the light-curve parameters measured from simulations versus real training data is shown in Figure \ref{fig:lcpar}. Spectra are also generated as part of these simulations as discussed in \citetalias{Kenworthy21} such that the total number of spectra, the distribution of their phases, and their noise properties as a function of wavelength for each individual survey are approximately equal to the real \citetalias{Kenworthy21} training sample.  %

\subsection{Varying the Input Training Data}
\label{subsec:trainingvariants}

We modify the input data used for SALT3 training in several ways, listed below, to explore the potential effect of removing less reliable data from the model training and the impact of the number and quality of the spectroscopic data on the resulting model surfaces.

\begin{itemize}
    \item {\bf Removing low-$z$ SNe without measured filter throughputs}.
    The low-$z$ sample used in \citetalias{Kenworthy21} is a compilation of data from various surveys dating back to the 1990s.  Prior to the CfA3 sample \citep{Hicken09}, 
    the filter transmission functions were estimates and color transformations were used so that synthetic colors matched observations of Landolt standard stars.
    The lack of precise filter transmissions can introduce unknown systematic uncertainties in the calibration of the SN photometry.  In particular, \citet{Scolnic15,Brout21Fragilistic} found that the photometry of these surveys could be systematically off by up to 3\%, but they lacked the statistics to correct for these offsets.  
    
    Thanks to the recent wealth of low-$z$ data from CSP, Foundation, and the later CfA surveys, it is possible to train the SALT3 model after removing data without measured filter transmissions: the Calan/Tololo and CfA1-2 low-$z$ samples. %
    Rather than testing the systematic effect of calibration offsets,we test the statistical impact of excluding these low-z samples from the training.
    All other simulated data for this variant remain the same as in our baseline simulations, although the total amount of photometric and spectroscopic data is significantly smaller due to removing these samples. On average, each random sample of this variant consists of 1000 SNe with 609 spectra. The loss of SNe is about $4\%$, while the loss of spectra is nearly $50\%$.

    \item {\bf Removing $U$-band data}.
    The observer-frame ultraviolet data ($U$/$u$ band, central wavelength $<$4000\AA) from low-$z$ surveys and SDSS are particularly difficult to calibrate, and in past analyses $U$-band model trainings have been found to be unreliable \citep{Kessler09}.  Recent efforts to improve the calibration of legacy SN\,Ia data \citep{Scolnic15,Currie20,Brout21Fragilistic} have not attempted to recalibrate the $U/u$ bands.  Because the SALT procedure can use well-calibrated high-$z$ data in the optical bands (with central wavelength $>4000$ \AA) to train the UV model, it might be preferable to omit these poorly calibrated data entirely. We therefore test the effect of removing the $U$/$u$ band data from our simulated training set.
    
    \item {\bf Reducing the number of spectra}.
    Including both photometric and spectroscopic data in the SALT3 training allows the model to more reliably account for variation in spectral features, improving the fidelity of $K$-corrections.  However, spectral data are expensive to obtain relative to photometric observations and must be iteratively recalibrated to match the best-fit model during the training process.  For this reason, 
    we test the effects of randomly removing half of the spectra from the simulated training data; this allows us to explore the degree to which spectroscopic data can be removed from the training process while still yielding reliable SN\,Ia distances.
    
    \item {\bf Including spectroscopic calibration errors}.
    The relative, wavelength-dependent flux calibration of SN spectroscopy can be highly uncertain.
    For this reason, {\tt SALTShaker} re-calibrates the spectra to match the best-fit SALT model during each iteration of the training process. We test the robustness of this recalibration procedure by including simulations of mis-calibrated spectra in the training data. For each simulated spectra, a multiplicative calibration warp in the form of $dF/d\lambda \rightarrow dF/d\lambda \times (1+s\lambda)$ is applied, with $s$ being randomly selected between $-10^{-5}$ \AA$^{-1}$ and $+10^{-5}$ \AA$^{-1}$.  This degree of re-calibration changes the spectral flux by up to $\pm$6\% over the range from 2800 to 8700~\AA, which are the minimum and maximum effective wavelengths of the photometric filters used in the training.  The {\tt SALTShaker} spectral recalibration procedure uses a polynomial to estimate the spectral warping; however, unlike the warping used in the simulations, the spectra are recalibrated using the exponential of a polynomial (a different functional form from the simulations), with coefficients estimated during the training process \citepalias[their Equation 6]{Kenworthy21}.
    
    \item {\bf Including host-galaxy contamination in the spectra}.
    In many galaxies, particularly at higher redshifts, the brightness of a SN\,Ia is often comparable to the surface brightness of its host galaxy.  There can be large uncertainties in subtracting host-galaxy light from the SN spectrum, which could add unphysical spectral features to the SN model or even shift the model's color if the spectral re-calibration procedure is insufficient to remove higher-order calibration offsets.  In the real \citetalias{Kenworthy21} input data, host-galaxy emission lines were removed in the low-$z$ data, and contamination was removed in the high-$z$ data by fitting for a simultaneous SN and host spectrum, and subtracting the host component \citep{Guy07}.  However, this procedure could leave some fraction of residual host contamination.
    
    We simulate the host-galaxy contamination effect by adding a scaled host-galaxy spectrum to the SN spectra, with the host-galaxy spectrum randomly selected from a small host-galaxy library.  This host-galaxy library is compiled from a random subset of high-S/N spectra of the host galaxies of SNe discovered during the Lick Observatory Supernova Search \citep{Filippenko01}. The host-galaxy spectrum is scaled such that
    \begin{equation}
        \mathrm{HOSTSNFRAC} = \frac{S_{host}\int d\lambda[dF_{host}/d\lambda]}{\int d\lambda [dF_{peak}/d\lambda]}.
    \end{equation}
    where $dF_{host}/d\lambda$ is the host spectra, $dF_{peak}/d\lambda$ is the SN spectrum at peak brightness, and $S_{host}$ is the normalization factor for the host spectrum for a given host-galaxy contamination fraction (HOSTSNFRAC).
    We test HOSTSNFRAC of $100\%$, $50\%$ and $10\%$, with respect to the SN brightness at peak. Example simulated SN spectra with different fractions of host contamination and their simulated host-galaxy spectra are shown in Figure \ref{fig:hostspec}.

\end{itemize}

\begin{figure*}[htb!]
\centering
\includegraphics[width=0.9\textwidth]{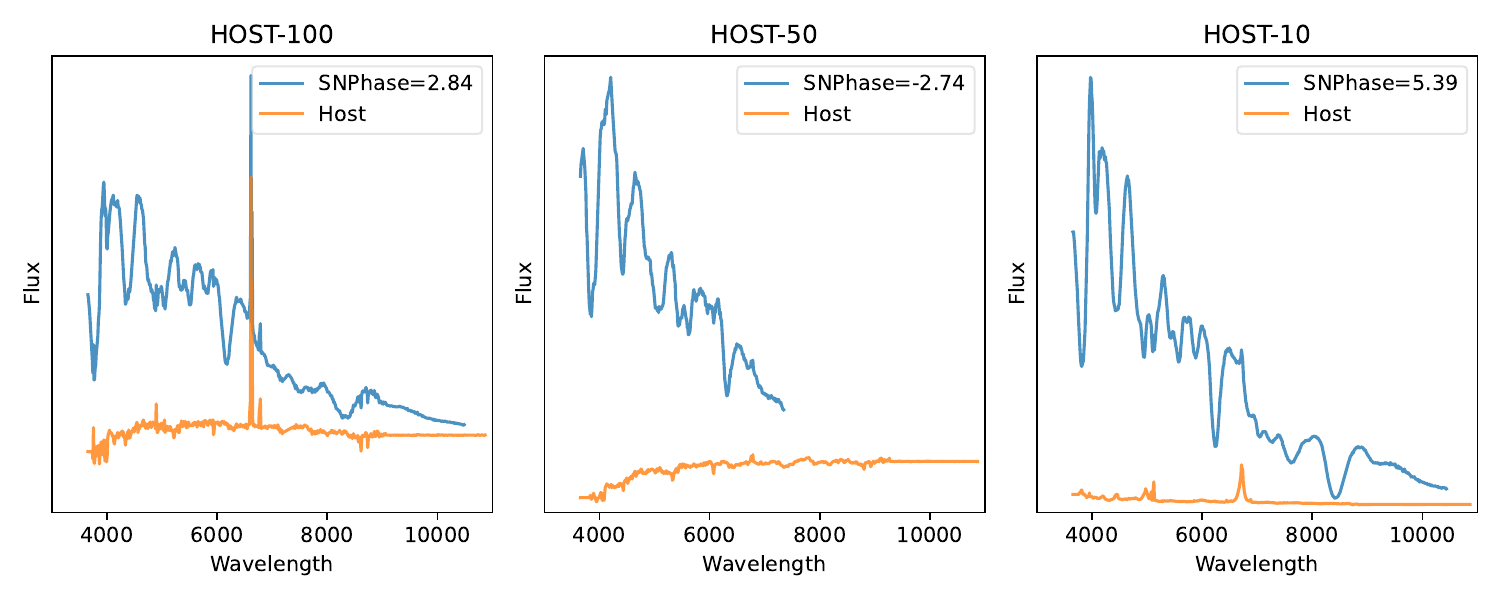}
\caption{Examples of simulated SN and host-galaxy spectra. Blue: simulated final SN spectra with host-galaxy contaminations added; from left to right the host-galaxy fractions are 100\%, 50\% and 10\% relative to the SN peak brightness. Orange: the simulated host-galaxy spectra for each SN, scaled relative to the SN peak brightness. The uncontaminated SN spectra are simply a subtraction between the final SN spectra and the host-galaxy spectra.}
\label{fig:hostspec}
\end{figure*}

\subsection{Varying the Input SN SED with the BYOSED Model}\label{subsec:byosed_variants}
The SALT formalism models the SN\,Ia SED with only one shape and one color parameter, but additional variability in SN\,Ia spectra and photometry has been seen \citep[e.g.,][]{Fakhouri15,Saunders18,Leget20,Boone21}.
Observed correlations between Hubble residuals and host-galaxy mass \citep{Kelly10,Lampeitl10,Sullivan10}, other host properties \citep{Rigault13,Jones18,Roman18,Rigault20,Kelsey21}, host-galaxy reddening \citep{Brout21,Rose22,Meldorf22}, and the potential correlation between Hubble residuals and SN ejecta velocity \citep{Siebert19,Dettman21}, give additional evidence for variability beyond the standard SALT parameters.

BYOSED enables us to test the reliability of the SALT formalism for measuring accurate distances from a flexible suite of SN\,Ia models.  These models are not necessarily intended to be true representations of real SN\,Ia data, but instead are {\it a priori} plausible models for the ways in which SN\,Ia spectra could depend on the SN shape, color, and additional standardization parameters.  We use the following BYOSED model variants to test the {\tt SALTshaker} training framework:

\begin{itemize}
    \item {\bf Shape and Color}.
    We follow \citet{Pierel21} in simulating a baseline \citet{Hsiao07} model with shape variations modeled as a simple wavelength-independent time dilation and color variations modeled using the SALT2 color law. The \citet{Hsiao07} model presents an average spectrum at every phase. While this model uses the same color law as the SALT2 model, it does not have the SALT spline-basis interpolation features imprinted on the model surface. %
    \item {\bf Spectral variation as a function of host-galaxy mass}.
    Following \citet{Pierel21}, we add simulated host galaxy-based variation to the shape$+$color simulations above using the same host-galaxy mass perturbers created by \citet{Pierel21}. \citet{Pierel21} generates the perturbers by making composite spectra from SNe with host-galaxy masses ${\rm log(M_{\ast}/M_{\odot}) > 10}$ and ${\rm log(M_{\ast}/M_{\odot}) < 10}$ from the Kaepora database \citep{Siebert19}.  
    Though some of the observed spectral variation is likely due to differences in the distribution of $x_1$ and $c$ between high- and low-mass host galaxies, these composites are an approximation for testing how effectively a SN that varies as a function of its host-galaxy properties can be modeled by a procedure that does not include host-galaxy dependence in its model assumptions.  Fifty percent of our simulated SNe use the low-mass composite versus the high-mass composite.
    We simulated two different host-galaxy mass perturbers as in \citet{Pierel21}, one with a static host-galaxy component with no redshift evolution, the other with a host-galaxy component that decreases in amplitude as a function of redshift. 
    
    \item {\bf Spectral variation as a function of SN velocity}.
    Similar to the host-galaxy mass procedure above, we added velocity perturbers constructed by \citet{Pierel21}. These velocity perturbers are generated by constructing spectral composites for SNe with low and high \ion{Si}{2} velocities (the divide between low and high velocity is located at the sample median of 11,000~km/s).  We draw from the observed distribution of \ion{Si}{2} velocities following \citet{Pierel21} to randomly assign velocities to SNe.  This variant is meant to mimic the spectral variations that caused a 2.7-$\sigma$ shift in Hubble residual as a function of \ion{Si}{2} velocity found by \citet{Siebert19}.
\end{itemize}

\section{Results}
\label{sec:results}

In this section, we compare results from each of the SALT3 training variants presented in the previous section. We first run our pipeline (Section \ref{sec:pipeline}) 40 times for the baseline scenario described in Section \ref{subsec:baseline_sim}; the baseline simulations were designed to reproduce the SALT3.K21 training sample. Then for each training variant described in Section \ref{subsec:trainingvariants} and Section \ref{subsec:byosed_variants}, we run our pipeline 20 times. Each pipeline run starts with a different random seed. We compute the average of the trained model surfaces, inferred distances, nuisance parameters, and measurements of $w$ for the 40 baseline runs and the 20 runs of each variant. We compare the averages of each 20 variant results to the average of the 40 baseline results.

First, however, we examined $w$ results from the baseline simulations and found a surprising $w$ bias of $-0.024 \pm 0.006$ relative to $\Lambda$CDM.  We traced this bias to low-$z$ simulations that adopted the \citet{Scolnic16} $x_1$ and $c$ distributions for the legacy low-$z$ data, which is the sample containing most of our spectra and some of the  best-sampled photometry.  These distributions do not fully match those of the \citetalias{Kenworthy21} training sample and in particular the color distribution of the \citetalias{Kenworthy21} data is wider; when simulating the low-$z$ training sample with $x_1$ and $c$ distributions that more closely match those of \citetalias{Kenworthy21}, we find a statistically insignificant $w$ bias of $w=-1.005 \pm 0.009$.  We discuss the potential reasons for the $w$ bias in the baseline version of the simulations in Appendix \ref{ap:baseline_sim} and conclude that training sets consisting of SNe with a wider distribution of colors, as in \citetalias{Kenworthy21}, will be more robust to cosmological biases.

For simulations that use the BYOSED models as the underlying models, we only compare the distances, and the measurements in $w$, since it is not meaningful to compare the model surfaces when the underlying models used for simulation are different. We compare the BYOSED variants with the BYOSED stretch and color model described in \ref{subsec:byosed_variants}, which serves as the baseline for the BYOSED models. The simulations that use the baseline BYOSED model yield $w=0.971\pm0.016$, which is consistent with an expected offset found in \citet{Pierel21}.

For convenience, we define short names for each training variant in Table \ref{table:shortnames}.  Below, we discuss variations in the trained model surfaces and the color law (Section \ref{sec:modelvar}), changes in the resulting distance moduli (Section \ref{sec:distancevar}) and changes in the nuisance parameters and cosmological parameters (Section \ref{sec:cosmovar}).

\begin{table*}[htb!]
    \centering
    \caption{Short names for each training variant}
    \begin{tabularx}{0.9\textwidth}{ l X }
    \hline\hline
    \bf Variant     &  \bf Description \\
    \hline
    \nolowz{}     & Removing low-$z$ SNe observed prior to the CfA3 sample, which lack measured filter throughputs\\
    \nou{}        & Removing $U$/$u$ band data, which are difficult to calibrate\\
    \miscal{} & Adding color-dependent calibration offsets to the simulated spectra\\
    \halfspec{}   & Removing (randomly) half of the input spectra for training\\
    \hline
    \hosthundred{}    & Including host-galaxy contamination with a fraction as bright as 100\% of the SN peak brightness\\
    \hostfifty{}     & Including host-galaxy contamination with a fraction as bright as 50\% of the SN peak brightness\\
    \hostten{}      & Including host-galaxy contamination with a fraction as bright as 10\% of the SN peak brightness\\
    \hline
    \byobase{} & Baseline BYOSED model with stretch and color effects\\
    \byovel{}  & BYOSED model with a velocity effect added to the baseline\\
    \byohost{} & BYOSED model with a static host-galaxy mass effect added to the baseline\\
    \byohostz{}    & BYOSED model with a redshift dependent host-galaxy mass effect added to the baseline\\
    \hline
    \end{tabularx}
    \label{table:shortnames}
\end{table*}

\subsection{Variations in the trained model}
\label{sec:modelvar}

In this section, we examine the differences in the trained SALT3 model surfaces given the training variants described in Section \ref{sec:data}, including the principal components $M_0$ and $M_1$ and the color law $CL$.

\subsubsection{Spectral components}
\label{sec:spectralcomp}

We first examine the changes in model components with respect to the baseline $M_0$ component:
\begin{align*}
 \Delta M_0/M_{0, {\rm baseline}} &= (M_{0, {\rm variant}} - M_{0, {\rm baseline}})/M_{0, {\rm baseline}} \\
 \Delta M_1/M_{0, {\rm baseline}} &= (M_{1, {\rm variant}} - M_{1, {\rm baseline}})/M_{0, {\rm baseline}}. \numberthis
\end{align*}

\noindent We note that because $M_1$ can be zero or negative, we define it fractionally with respect to $M_0$.  Therefore, $\Delta M_1/M_{0, {\rm baseline}}$ is the fractional difference in the predicted light curve or spectrum of a SN having an $x_1$ parameter that is one standard deviation away from the mean.

\begin{figure*}[htb!]
\centering
\includegraphics[trim=1cm 0.1cm 1cm 0.1cm,width=\textwidth]{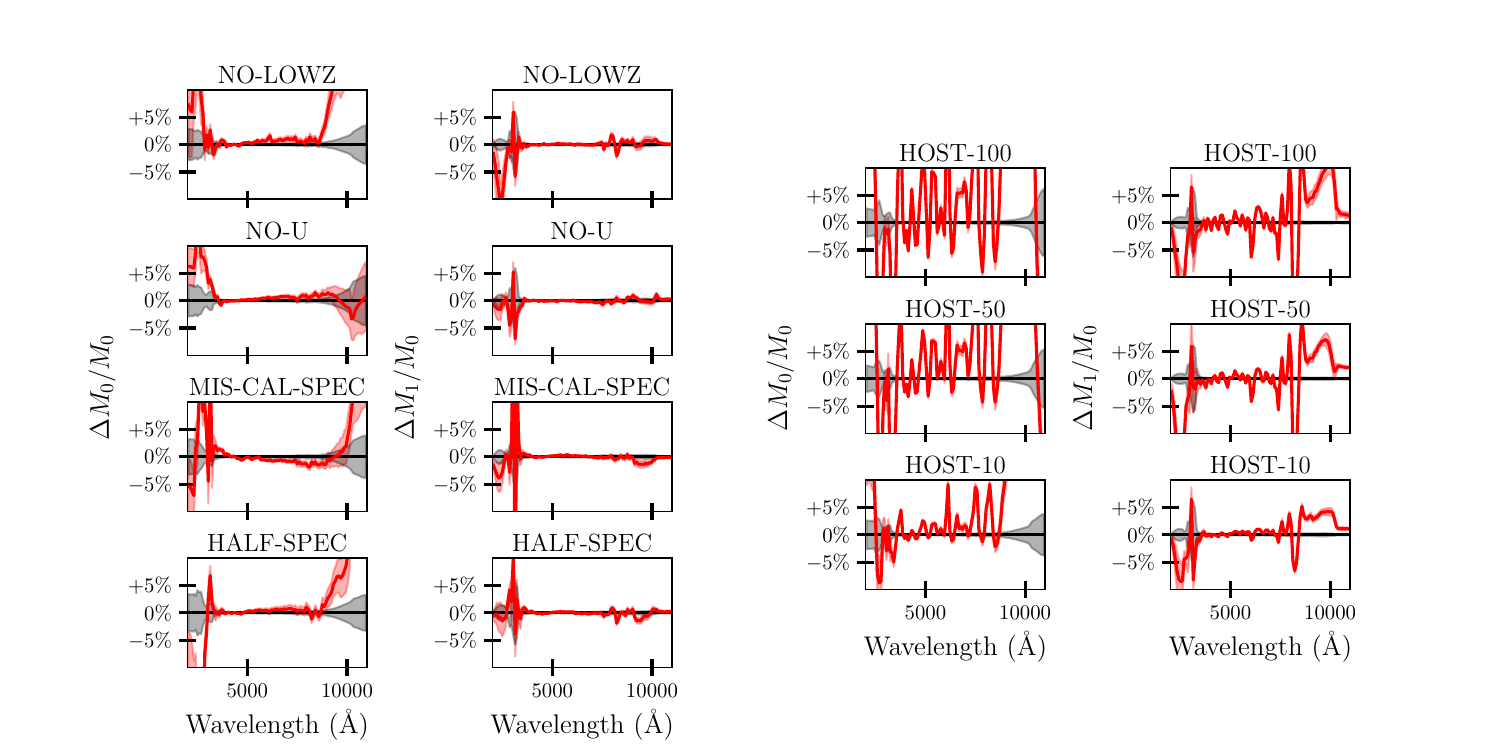}
\caption{Relative changes of model components $M_0$ and $M_1$ at peak for different variants of the training data with respect to the baseline $M_0$ component, with the gray area showing the error of the mean of the baseline $M_0$ component. Left: model changes for the following training data variations: removing the legacy low-$z$ data, removing the $u/U$ bands, adding calibration offsets to the training spectra, and removing 50\% of the spectral data. Right: model changes after including 100\%, 50\% and 10\% host-galaxy contamination in the spectral data, with contamination scaled relative to the SN brightness at maximum light.}
\label{fig:specvariants}
\end{figure*}

Figure \ref{fig:specvariants} shows the relative model changes in wavelength space for the SALT3 model components at SN peak brightness. For the variants shown in Figure \ref{fig:specvariants} (left), the $M_0$ components are consistent within 2$\%$ with the baseline model between 3000~\AA\ and 7500~\AA, but deviate more in the bluer and redder regions beyond that wavelength range, by up to $\sim 10\%$. The $M_1$ components are mostly consistent despite a larger variation in the bluer region below $3000$~\AA. For variants with different fractions of host-galaxy contaminations shown in Figure \ref{fig:specvariants} (right), the models are mostly consistent within 5\% with respect to the baseline when there is $10$\% host-galaxy light, but unsurprisingly, large ($>$5\%) spikes are seen in the $M_0$ components and some regions of the $M_1$ components when there is greater host-galaxy contamination.

\subsubsection{Integrated fluxes}
\label{sec:spectralcomp}

\begin{figure*}[htb!]
\centering
\includegraphics[trim=1cm 0.1cm 1cm 0.1cm,width=\textwidth]{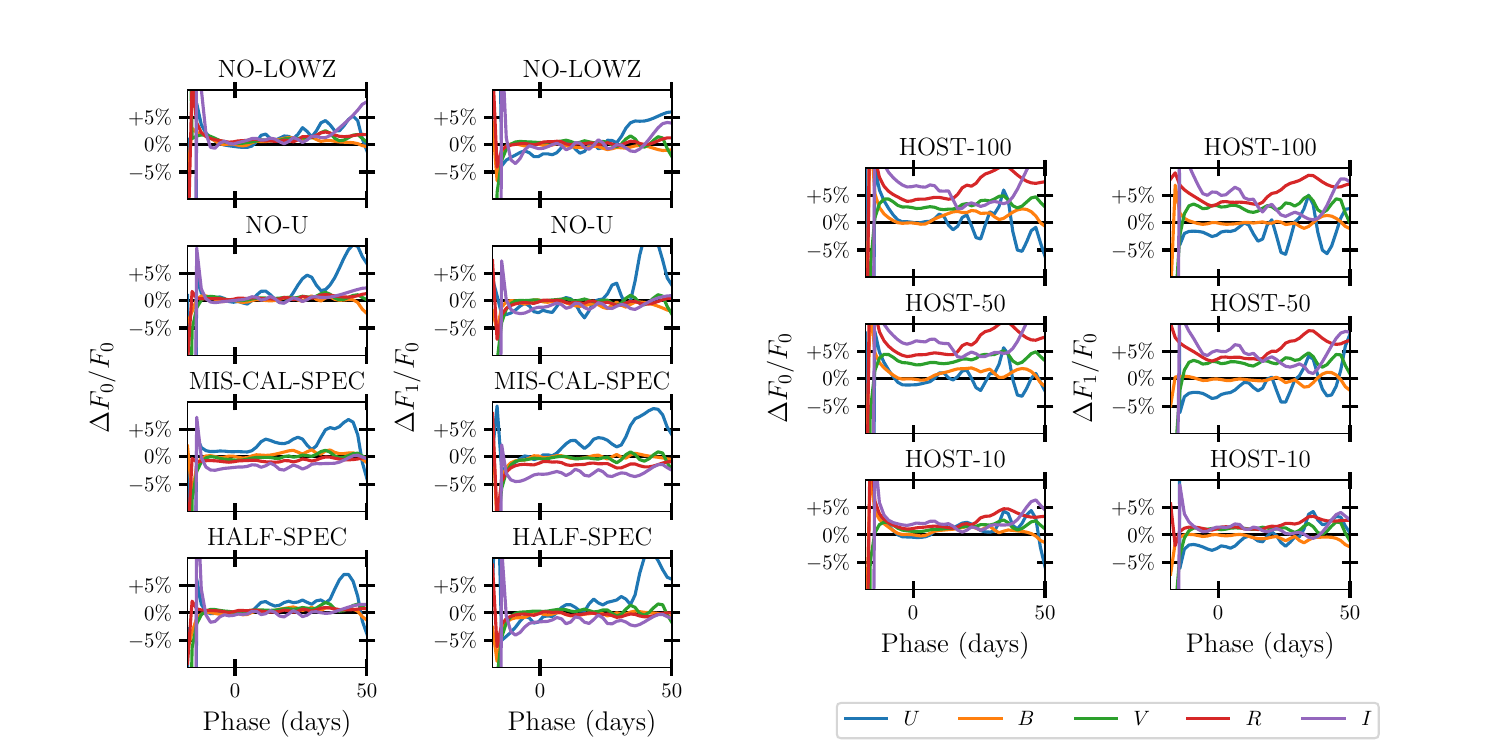}
\caption{Same as Figure \ref{fig:specvariants} but showing relative changes of integrated model fluxes $F_0$ and $F_1$ with respect to the baseline model flux $F_0$. Individual colors show the model components when integrated over the $UBVRI$ passbands.}
\label{fig:lcvariants}
\end{figure*}

In Figure \ref{fig:lcvariants}, we show the relative model flux variations in the $UBVRI$ bands. We integrate the model components $M_0$ and $M_1$ over the $UBVRI$ passbands for each variant, and show the relative changes with respect to the baseline $M_0$ model flux (also integrated over the $UBVRI$ passbands): 
\begin{align*}
 \Delta F_0/F_{0, {\rm baseline}} &= (F_{0, {\rm variant}} - F_{0, {\rm baseline}})/F_{0, {\rm baseline}} \\
 \Delta F_1/F_{0, {\rm baseline}} &= (F_{1, {\rm variant}} - F_{1, {\rm baseline}})/F_{0, {\rm baseline}}. \numberthis
\end{align*}

\noindent We find that the M$_0$ component for the $BVRI$ light curves is consistent to within 2\% at phases greater than $-15$~days for \nolowz{}, \nou{}, \miscal{} and \halfspec{}. %
The $M_1$ component shows larger variations, up to $\sim$3\% and biases of up to 4\% relative to $M_0$.

The $U$-band model fluxes show significantly greater variation compared to the other bands, however; the training variants \nolowz{}, \nou{}, \miscal{} and \halfspec{}{} all have substantially higher variation compared to the other bands in the UV model fluxes.  This is unsurprising, as each of these variants removes a significant fraction of the available $U$-band spectra or photometry.  This test therefore indicates that for the $U$-band SALT model to be robust, additional observations are needed to cover the rest-frame $U$ band, likely with corresponding high-S/N spectra, perhaps from future facilities that will have well-calibrated $u$-band data such as the Rubin Observatory.  Furthermore, the increased variation in $\Delta M_1$ indicates that there is value in obtaining a larger SALT3 training set in order to better constrain the first principal component of the model. Finally, when the mis-calibration effect is simulated in the SN spectra, we see $\sim$2-5\% offsets in $U$ and $I$, which are near the blue and red ends of the model's spectral range, respectively, and
which may be a result of limitations in the SALT3 re-calibration procedure.

In Figure \ref{fig:lcvariants} (right) we examine the effect of including host-galaxy contamination in the SALT training spectra in more detail by showing model flux variations from the 100\%, 50\%, and 10\% contamination training variants.  While \hostten{} has variation at the level of $\sim$3\% across most of the phase range, higher contamination yields further degradation in the training when using SN spectra with strong host-galaxy contamination. 
The \hostfifty{} and \hosthundred{} variants show color-dependent offsets of $\gtrsim$5\% in the $U$, $R$, and $I$ band model fluxes; we note that because the model is normalized to maximum light in the rest-frame $B$-band, this band shows smaller variation at the few-percent level.  High-$z$ SNe in particular tend to be fainter relative to the local surface brightness of their host galaxies, which can result in substantial host contamination; this level of bias in the model surfaces shows that high-$z$ SN spectra must have their host-galaxy contributions carefully removed to avoid model biases.

\subsubsection{Color law}

\begin{figure}[htb!]
\centering
\includegraphics[trim=0.5cm 0.1cm 0.2cm -0.5cm, width=0.45\textwidth]{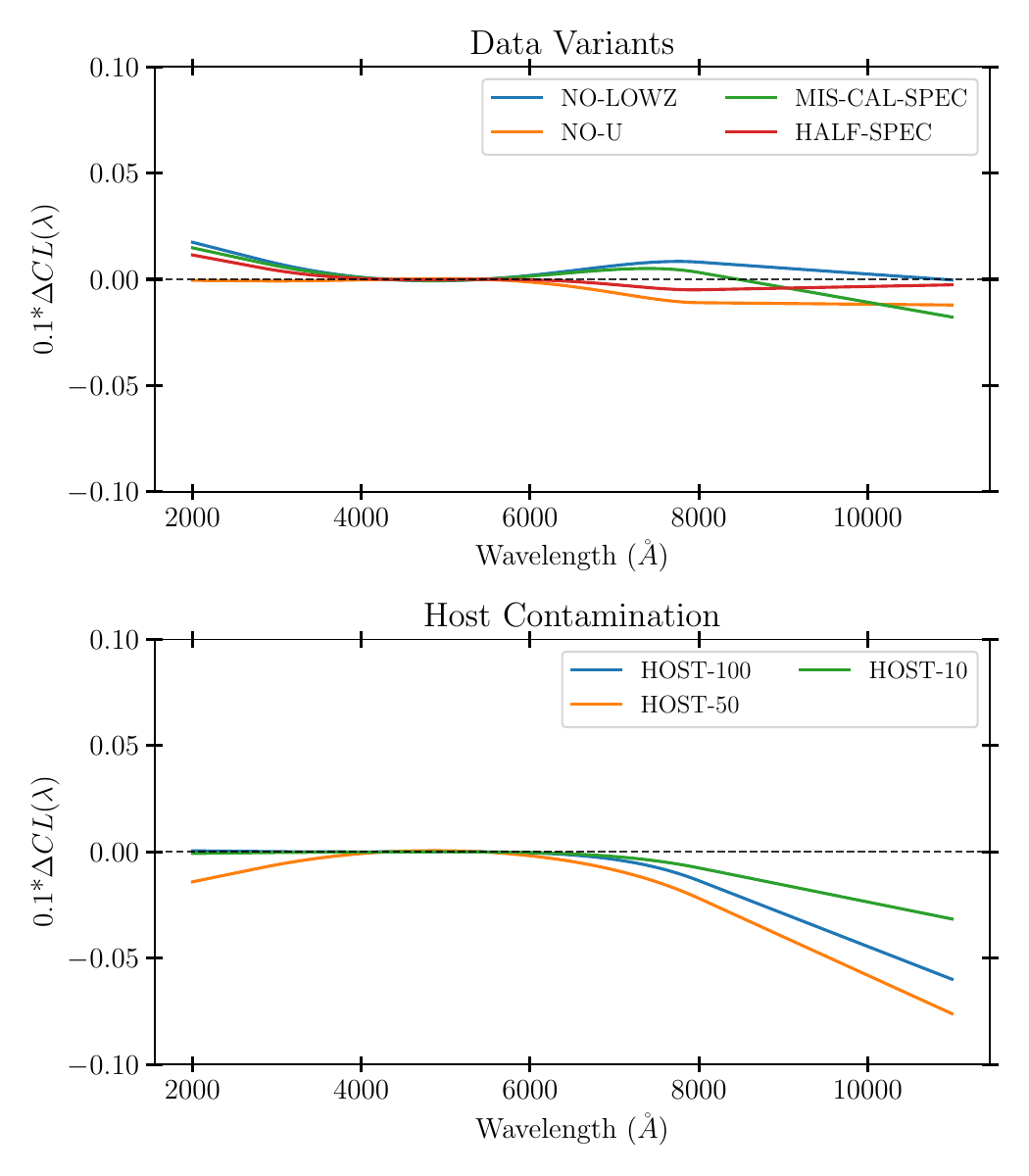}
\caption{Difference in color law for each training data variation with respect to the baseline. The coefficient 0.1 is chosen to show the change in predicted magnitude as a function of wavelength for a SN with $c=0.1$.}
\label{fig:colorlaw}
\end{figure}

The change in color laws of each data %
variant are shown in Figure \ref{fig:colorlaw}. We describe the color law change with respect to the baseline with the following equation: 

\begin{equation}
\Delta m(c,\lambda) = c \times ( CL(\lambda)_{variant}- CL(\lambda)_{baseline})
\end{equation}

\noindent Here, $\Delta m(c,\lambda)$ is the change in magnitude as a function of wavelength for a SN with color $c$ relative to the baseline model.  We choose a nominal $c = 0.1$ for the comparisons in Figure \ref{fig:colorlaw}, equal to a difference in reddening between the $B$ and $V$ bands, compared to an average SN\,Ia, of 0.1~mag.  Typical SN\,Ia cosmology cuts limit $c < 0.3$, so we note that the reddest --- and the bluest --- SNe\,Ia in a typical data set would be biased by three times the values shown in this figure.

Similarly to the spectral components, the color laws are consistent to 2\% or better for $c = 0.1$ between 3000\AA\ and 7500\AA, but have larger deviations in the bluer and redder regions.  The SALT3 color law is extrapolated beyond 8000\AA, and these extrapolated regions at redder wavelengths in particular can deviate by up to $\sim$10\% when including host contamination effects, and therefore should not be considered robust to reasonable variations in the training data.

\subsection{Differences in distance moduli}
\label{sec:distancevar}

\begin{figure*}[htb!]
\centering
\includegraphics[trim=0.2cm 0.2cm 1cm 0.2cm,width=0.9\textwidth]{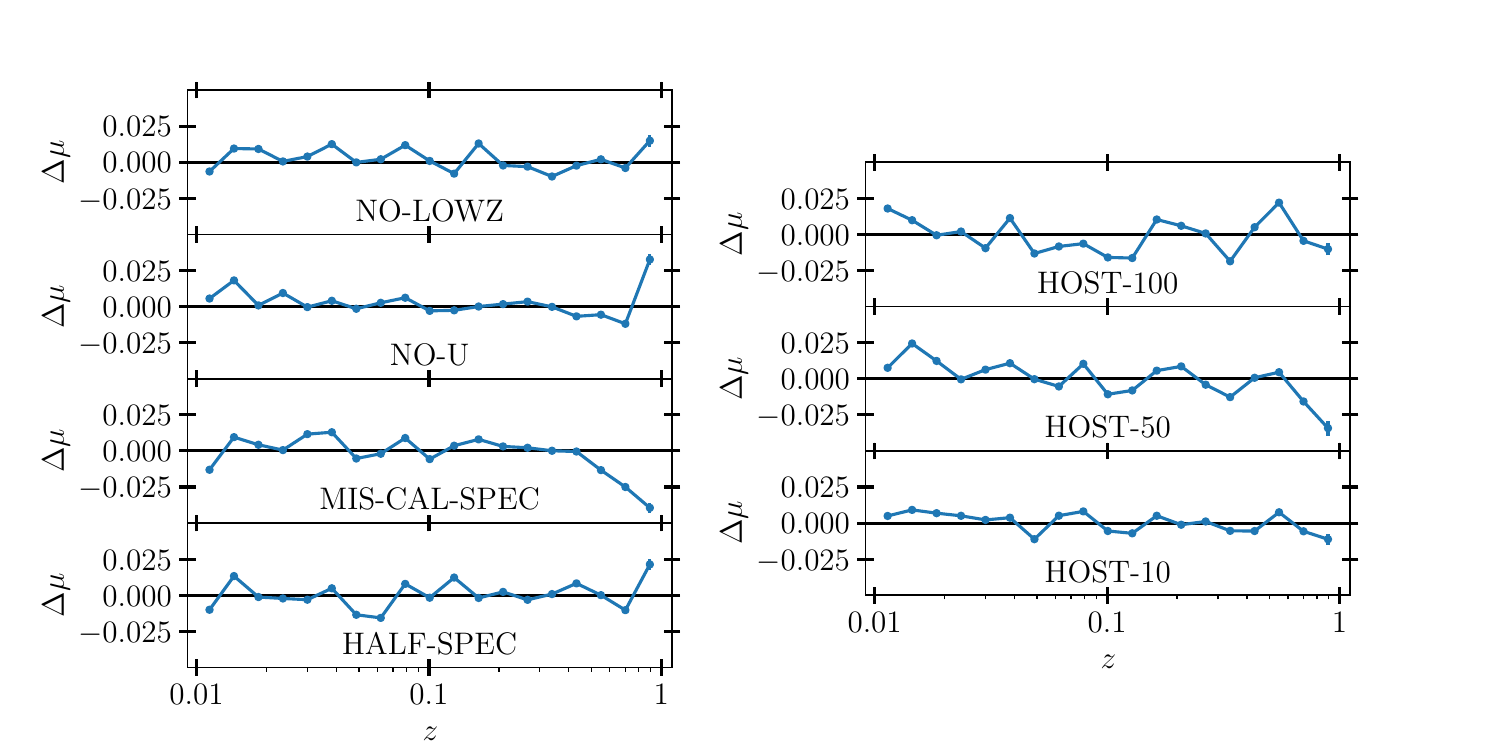}
\caption{Changes in the distance moduli as a function of redshift for each training variant, with respect to the baseline.  Left: distance modulus changes for the following training data variations (from top to bottom): removing the legacy low-$z$ data, removing the $u/U$ bands, including mis-calibrated spectra, and removing 50\% of the spectral data. Right: distance modulus changes when including 100\%, 50\% and 10\% host-galaxy contamination, with respect to SN brightness at maximum light, in the input spectral data.}
\label{fig:hubblediag1}
\end{figure*}

\begin{figure}[htb!]
\centering
\includegraphics[trim={0.5cm 1cm 0.2cm 1.5cm},clip,width=0.45\textwidth]{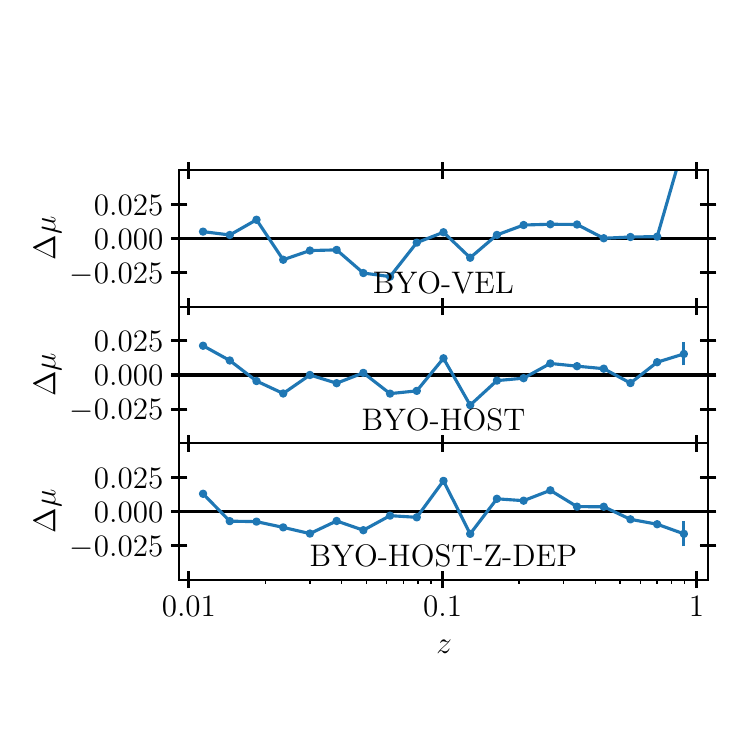}
\caption{Changes in the distance moduli (with respect to the BYO-STRETCH-COLOR model) as a function of redshift when simulating the input data using the BYOSED model. Top: BYOSED model with an ejecta velocity effect added. Middle: BYOSED model with a constant host-galaxy mass step effect added. Bottom: BYOSED model with a redshift dependent host-galaxy mass effect added.}
\label{fig:hubblediag2}
\end{figure}

We show the changes in distance moduli for each data variant with respect to the baseline in Figure \ref{fig:hubblediag1}, and for BYOSED simulations in Figure \ref{fig:hubblediag2}. For $z<0.6$, the distance modulus changes are below 0.025 for each of the input data variants. Larger changes of $\sim$0.5 are seen for $z>0.6$, with the largest changes coming from \miscal{} and \hostfifty{}.

The RMS of the Hubble residuals is consistent to within $\sim$2\% across all variants that are simulated with the extended SALT2 model, with the \miscal{} model having the largest scatter, 0.149~mag, but negligibly higher than the baseline value of 0.147~mag.
For the BYOSED variants, the RMS of the Hubble residuals are higher, $\sim$0.245.
The reduced $\chi^2$ of the SALT3 light curve fitting is also consistent across all the variants that are simulated with the extended SALT2 model.  By excluding the error term due to the in-sample variance --- the degree to which the SALT3 formalism does not fully encapsulate the intrinsic variability of the underlying SN data --- we compare reduced $\chi^2$ across training options.  The baseline model has a median reduced $\chi^2 = 0.95$, which could indicate slight under-regularization in the training procedure or very modest overestimation of the uncertainties.

Most training options yield a comparable reduced $\chi^2$ to the baseline case, but the variants \nou{}, \hostten{}, \hostfifty{} and \hosthundred{} yield a slightly higher reduced $\chi^2$, ranging from $0.98$ to $1.01$. \miscal{} has a smaller reduced $\chi^2$ = 0.92.  The BYOSED models have slightly larger median reduced $\chi^2$ $\sim 1.08$, perhaps a consequence of these models being semi-independent of the SALT model framework.

In spite of the increased $\chi^2$ from some variants, it is encouraging that the distances change very little on average.  Biases in distances relative to the baseline model, averaged in redshift bins of $0 < z < 0.2$ versus $0.4 < z < 0.6$, are shown in Table \ref{table:w}.  All biases, except \hosthundred{} or \miscal{}, are consistent with zero with an uncertainty of less than 0.01 mag, demonstrating that the training procedure remains effective at standardizing its training sample, even if the model surfaces themselves are shifting on the few-percent level. \hosthundred{} and \miscal{} have a bias $\sim 0.02$ and $-0.01$, respectively.

\subsection{Biases on cosmological and nuisance parameters}
\label{sec:cosmovar}

Despite the apparent shifts in $M_0$ and $M_1$ described in Section \ref{sec:spectralcomp}, we find that inferred distances are consistent. This may be due to the compensating shifts in $\alpha$ and $\beta$ in the distance estimation stage.
The differences in these nuisance parameters are shown in Table \ref{table:w}. 
We observe differences in $\alpha$ of $\lesssim$2$\sigma$ for the SALT2-extended model simulations except for one high-significance shift of 0.014 for the \hostten{} variant. %
Biases in the $\beta$ parameter are within $2$-sigma significance for all variants simulated with SALT2-extended, with a magnitude of up to 0.092.  %

Finally, Table \ref{table:w} shows the differences in measurements of $w$ from these different training variants.  All are consistent with the baseline model at the $\sim$2-$\sigma$ level ($\Delta w \lesssim 0.025$), but with the largest potential deviations coming from variants that adversely affect the reliability of the training spectra (including the \nolowz{} variant, which removes many spectra).  %

We compare the BYOSED models with BYOSED-STRETCH-COLOR, and show the differences in $w$ in Table \ref{table:w_byos}. All values of $w$ are consistent to within 0.025, showing that the SALT3 training procedure can largely account for the effect of possible correlations between the host galaxy and the SN\,Ia SED, or additional SN properties such as SN velocity. %

\begin{table*}[htb!]
    \caption{Biases on Nuisance Parameters and $w$ for SALT-based Variants}\label{table:w}
    \centering
    \begin{tabular}{lrrrr}
    \hline \hline
    Training Variant&$\Delta\mu^{\rm a}$&$\Delta\alpha$&$\Delta\beta$&$\Delta w$\\
    \hline

NO-LOWZ&$-0.003\pm0.005$&$-0.005\pm0.004$&$-0.056\pm0.033$&$0.019\pm0.008$\\
NO-U&$-0.008\pm0.006$&$-0.001\pm0.003$&$0.029\pm0.030$&$0.014\pm0.010$\\
MIS-CAL-SPEC&$-0.013\pm0.006$&$0.002\pm0.002$&$-0.029\pm0.031$&$0.025\pm0.013$\\
HALF-SPEC&$0.010\pm0.005$&$0.001\pm0.003$&$0.059\pm0.037$&$-0.007\pm0.012$\\
\hline
HOST-100&$0.021\pm0.005$&$-0.007\pm0.003$&$0.064\pm0.041$&$-0.011\pm0.009$\\
HOST-50&$0.004\pm0.006$&$-0.008\pm0.003$&$0.092\pm0.046$&$0.019\pm0.010$\\
HOST-10&$-0.001\pm0.005$&$-0.014\pm0.003$&$0.018\pm0.033$&$0.008\pm0.010$\\

\hline\\*[-2pt]

\multicolumn{5}{l}{
\begin{minipage}{0.65\textwidth}

$^{\rm a}$ Relative to the baseline fitting results, the difference between average Hubble residual at $0.01 < z < 0.2$ and the average Hubble residual at $0.4 < z < 0.6$.

\end{minipage}
}
\end{tabular}

\end{table*}

\begin{table}[htb!]
    \centering
    \caption{Biases on $w$ for BYOSED Variants}
    \begin{tabular}{lrr}
    \hline \hline
    Training Variants&$\Delta\mu$&$\Delta w$\\
    \hline

BYO-VEL&$0.004\pm0.008$&$-0.025\pm0.015$\\
BYO-HOST&$-0.001\pm0.008$&$-0.011\pm0.018$\\
BYO-HOST-Z-DEP&$0.001\pm0.008$&$-0.016\pm0.016$\\
\hline
    \end{tabular}
    \label{table:w_byos}
\end{table}

\section{Discussion}
\label{sec:discussion}

In this paper, 
we test the robustness of the newly developed {\tt SALTShaker} code and the SALT3 model, and quantify the systematic uncertainties of the training procedure on cosmological parameter measurements.  We explore the effect of removing legacy low-$z$ data without measured filter throughputs, observer-frame $u/U$ data (which is poorly calibrated in the real training data), and a large fraction of the spectroscopic training data.  We also test the consequences of mis-calibrated SN spectra and the effects of host-galaxy contamination in the spectra.  Finally, we use a simulated model independent of SALT (BYOSED; \citealp{Pierel21}) to test whether a SALT3 model trained on these data can produce consistent distances from $0 < z < 1$.

We generally find better than 2\% consistency of the model across these different training variants. The most significant changes in the model are seen in the $U$ band, in pre-maximum light phases, and in the redder and bluer edges of the model surfaces. Host-galaxy contamination at $10\%$ of the SN maximum brightness does not produce large changes in the trained model; however, host contamination at the level of $>50\%$ of the SN maximum brightness can produce larger than $5\%$ model variations and slightly increases the $\chi^2$ when the data are fit to the resulting SALT3 model.

We also see evidence that the 1990s-era legacy data and the observer frame $U$ band data are playing important roles in constraining the model surfaces and the color law.  It appears that individual SNe with photometry spanning the full available model wavelength range ($U/u$ to $I/z$) may also improve the fidelity of the model, and the existing model training sample may need to be expanded to fully replace these data.  We suggest compiling additional training samples, particularly at low redshift, from CSP and Rubin Observatory photometry that include $uBVgri$ and $ugriz$ (and perhaps $y$) data, respectively, for the same SNe; this will help to fully constrain the model surfaces and color law simultaneously and allow us to remove the less-reliable low-redshift training samples.  SN spectra also appear to be important for constraining the model surfaces in the SALT3 training process, particularly at the bluest and reddest ends of the wavelength range; when half of our spectral training sample are removed, we find model variations on the order of up to $\sim$5-10\%.

Although these different training options demonstrate the phases and wavelength ranges where the SALT3 training is less robust, we see consistent distance measurements across nearly the full redshift range, as well as consistent cosmological parameter measurements to within 2\% in most cases; the \miscal{} variant has the largest deviation of 0.025 at 1.9-$\sigma$ significance.  Provided that the cosmological parameter estimation is run on data drawn from the same intrinsic distribution as the training data --- i.e., it is important that the SALT3 model is re-trained on the data used in a given cosmology analysis --- we find that cosmological parameter measurements from the SALT3 model are robust at the 2\% level to most realistic variations in the SALT3 training process.  However, we suggest that additional attention must be paid to the way in which SALT3 re-calibrates spectra during the training process, and to the contamination of the high-redshift training spectra by host-galaxy light. Additionally, we find that the SALT3 training process is sensitive to the color distribution of the input training data, and the resulting $w$ measurement can be biased by $\sim 2\%$ if the color distribution is not sufficiently wide.

Finally, we suggest using the pipeline developed in this work, or similar approaches, to propagate variants like these into the systematic uncertainty budgets of future cosmological analyses.  We can only be confident in our constraints on dark energy if we fully understand the assumptions that are propagated into SN standardization modeling.  Fortunately, our analysis indicates that  these types of systematic uncertainties in the training process bias $w$ at a level below the precision of current analyses. However, as hundreds of thousands of additional SNe\,Ia are discovered in the coming years, $w$ measurements become more precise, and these new data simultaneously improve our models for estimating SN distances and extend the viable wavelength range at which measuring SN\,Ia distances is possible, these types of end-to-end systematic uncertainty tests will become increasingly important.

\begin{acknowledgments}
M.D. is supported by the Horizon Fellowship at the Johns Hopkins University. Support for D.O.J. was provided by NASA through the NASA Hubble Fellowship grant HF2-51462.001 awarded by the Space Telescope Science Institute, which is operated by the Association of Universities for Research in Astronomy, Inc., for NASA, under contract NAS5-26555. Research on systematic uncertainties in supernova cosmology at Rutgers University is funded by DOE awards DE-SC0011636 and DE-SC0010008. The UC Santa Cruz team is supported in part by NASA grants 14-WPS14-0048, NNG16PJ34C, and NNG17PX03C, NSF grants AST-1518052 and AST-1815935, the Gordon and Betty Moore Foundation, the Heising-Simons Foundation, and by a fellowship from the David and Lucile Packard Foundation to R.J.F.
This work was completed in part with resources provided by the University of Chicago’s Research Computing Center.

\end{acknowledgments}
\vspace{5mm}

\software{Astropy \citep{astropy:2013, astropy:2018, astropy:2022}, 
Matplotlib \citep{Matplotlib:2007}, 
NumPy \citep{Numpy:2020}, 
pandas \citep{Pandas:2020, Pandas:2010},
SNANA \citep{Kessler09b},
SNCosmo \citep{sncosmo}}

\appendix
\counterwithin{figure}{section}

\section{Simulation details}
\label{ap:sim_pars}

We list the Asymmetric Gaussian parameters that are used to generate our simulations in Table \ref{table:x1c}.

\setcounter{table}{0}
\renewcommand{\thetable}{A\arabic{table}}
\begin{table}[h]
    \centering
    \caption{Asymmetric Gaussian parameters for $x_1$ and c populations }
    \begin{tabular}{ccccccc}
    \hline
        Survey & $\bar{x_1}$ & $\sigma_{x_1}^{+}$ & $\sigma_{x_1}^{-}$ & $\bar{c}$ & $\sigma_c^{+}$ & $\sigma_c^{-}$ \\
        \hline
        Low-z & 0.419 & 3.024 & 0.742 & -0.069 & 0.003 & 0.148 \\ 
        SDSS & 1.142 & 1.652 & 0.104 & -0.061 & 0.023 & 0.083 \\ 
        PS1 & 0.589 & 1.026 & 0.381 & -0.103 & 0.003 & 0.129 \\
        SNLS & 0.974 & 1.236 & 0.283 & -0.112 & 0.003 & 0.144\\
        DES & 0.973 & 1.472 & 0.222 & -0.054 & 0.043 & 0.101 \\
        Foundation & 0.703 & 1.0 & 0.5 & -0.068 & 0.033 & 0.125\\
        \hline\\*[-10pt]
        \multicolumn{7}{c}{
        \begin{minipage}{0.45\textwidth}
                \footnotesize Note: Following \citealp{Scolnic16}, $\bar{x_1}$/$\bar{c}$ is the value with the maximum probability for the asymmetric gaussian $x_1$/$c$ distribution, $\sigma^+$ and $\sigma^-$ are the corresponding gaussian widths on the low and high sides.
        \end{minipage}
        }
    \end{tabular}
    \label{table:x1c}
\end{table}

\section{Biases in $w$ from the Baseline Simulations}
\label{ap:baseline_sim}

Our baseline simulations described in Section \ref{subsec:baseline_sim} (which use $x_1$/$c$ populations derived by \citealt{Scolnic16}) yield a $w$ bias of $-0.024 \pm 0.006$. In order to investigate the origin of this bias, we create another set of baseline training set simulations following \citetalias{Kenworthy21}, which uses the $x_1$ and $c$ parameter values from the actual \citetalias{Kenworthy21} data sample.  For this sample, we find a statistically insignificant $w$ bias of $w=-1.005 \pm 0.009$ (averaged on 20 random samples). 

\begin{figure}[htb!]
\centering
\includegraphics[width=0.45\textwidth]{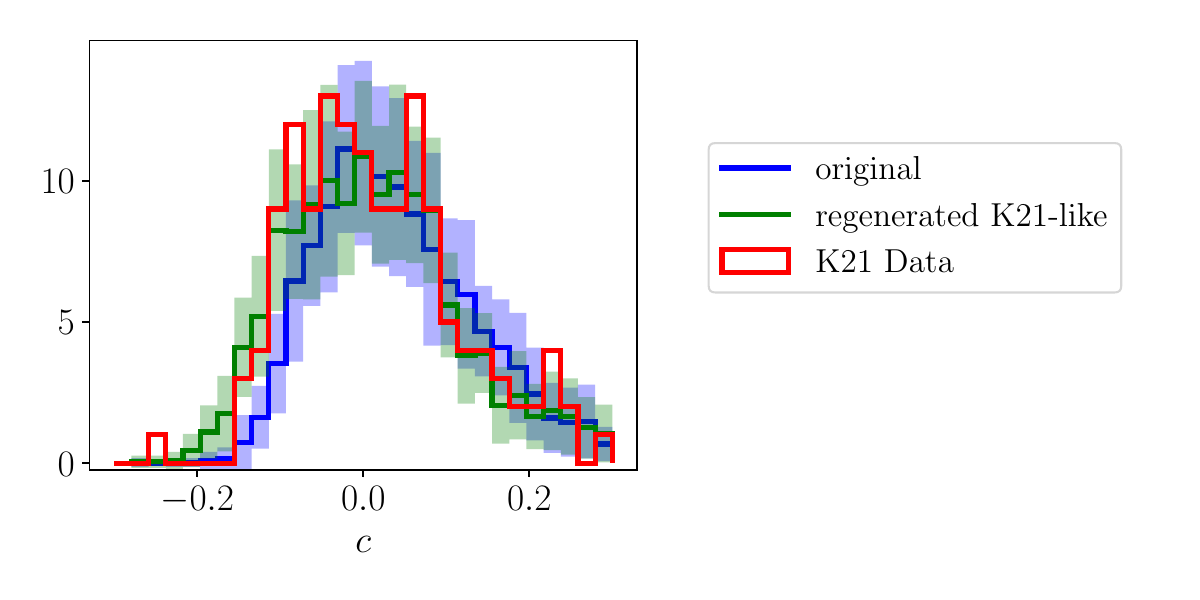}
\caption{Color distributions of the low-z training data from the original baseline simulations (with low-z $x_1/c$ parameters drawn randomly from previously generated populations, blue), the regenerated K21-like simulations (with K21 low-z $x_1/c$ parameters, green), and the K21 data (red). The shaded areas correspond to the standard deviation of the number in each bin for the multiple simulations.}
\label{fig:colordiff}
\end{figure}

The above change was applied only to the legacy low-$z$ simulations because we found that simulations based on the \citet{Scolnic16} $x_1$/$c$ populations did not fully match the \citetalias{Kenworthy21} data (Figure \ref{fig:colordiff}).  We note that this is not a deficiency of the \citet{Scolnic16} results; rather, the \citetalias{Kenworthy21} training data includes additional SNe that were not considered in \citet{Scolnic16}.  These legacy low-$z$ data, while a limited subset of the training, contain a majority of the spectra used in the training and also constitute some of the highest-S/N and best-sampled photometry.

Figure \ref{fig:colordiff} shows the color distributions of the low-$z$ training data from our original baseline samples, the regenerated \citetalias{Kenworthy21}-like baseline samples, and the \citetalias{Kenworthy21} data.  We find that the \citetalias{Kenworthy21}-like simulations (and the \citetalias{Kenworthy21} data) have significantly more SNe with blue colors compared to the original simulations, likely due to the addition of new data from the CfA4 survey, the CSP survey, and $z < 0.01$ SNe that are too nearby to be suitable for dark energy measurements (but can be used for light-curve training).

An alternative source of bias could be differences between $x_1$ and $c$ populations in the bias correction simulations versus the ``test" simulations.  Because the SALT training process enforces definitions of mean $x_1,c = 0,0$, subtle shifts in the simulated shape/color parameters of the simulated sample can cause mismatches between the training sample and our adopted $x_1/c$ distributions for the bias correction simulations.
However, slight changes in the mean of the $c$ distribution ($\sim$0.01) to make up for these effects had no significant effect on the $w$ bias.
We therefore infer that the difference in $w$ stems from differences in the two trained SALT3 models themselves, and that the trained models are sensitive to the color distributions of the input training data, though in future work we will re-compute $x_1$ and $c$ populations as part of our pipeline for each individual data set following the method of \citet{Scolnic16}. 

\begin{figure}[htb!]
\centering
\includegraphics[trim=0.2cm 0.2cm 0.1cm 0cm,width=0.45\textwidth]{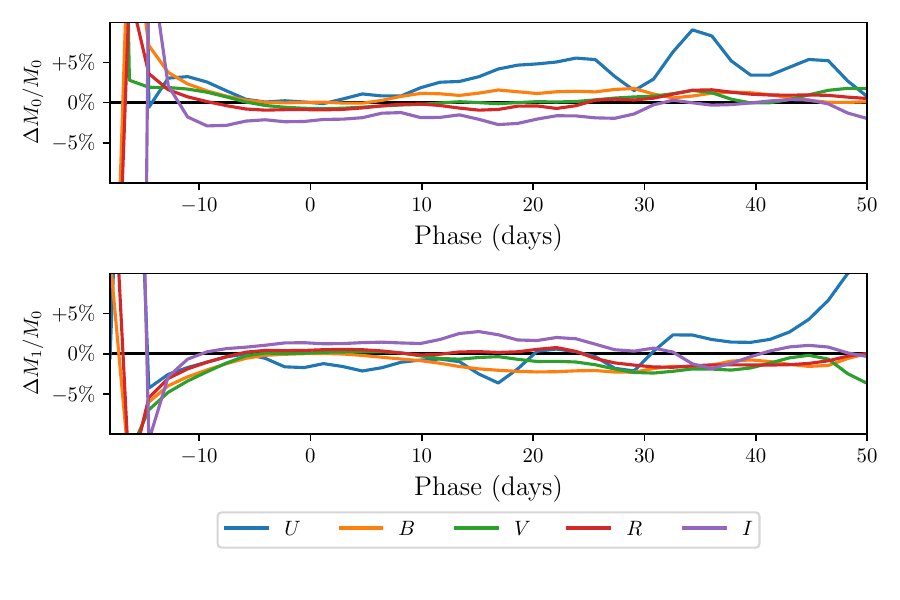}
\caption{Difference in the averaged SALT3 model flux (integrated over UBVRI bands) between the two samples described in Appendix \ref{ap:baseline_sim} --- the original baseline samples and the regenerated K21-like samples --- relative to the M$_0$ flux of the original baseline samples.}
\label{fig:seddiff}
\end{figure}

Fig \ref{fig:seddiff} shows the differences in the averaged trained model components trained with these two samples. We find a $>5\%$ difference in the rest-frame U band, and a slight difference in the color law. The color laws trained from these two samples are similar, with a slight difference for $\lambda > 7000 \AA$. We note that although on average the color laws are not significantly different between these two sets of simulations, we find that the SALT3 color law is particularly sensitive to the $c$ range of the training data for each individual random sample.
We advise that future model trainings use a training sample with a wide distribution of colors (similar to \citetalias{Kenworthy21}) in order to avoid subtle cosmological biases such as the one found here.

\bibliography{main}
\bibliographystyle{aasjournal}

\end{document}